\documentclass[prd ]{revtex4}%
\usepackage{caption}
\usepackage{amsmath}
\usepackage{graphicx}
\usepackage{epsfig}
\usepackage{dcolumn}
\usepackage{bm}
\usepackage{slashed}
\usepackage{amsfonts}
\usepackage{amssymb}%
\providecommand{\U}[1]{\protect\rule{.1in}{.1in}}
\begin{document}
\title{Can the symmetry breaking in the SM be determined \\by the "second minimum" of the Higgs potential? \bigskip}
\author{Alejandro Cabo$^{\ast},$ Jose Carlos Suarez$^{\ast\ast}$,\\Denys
Arrebato$^{\ast\ast\ast}$, Fernando Guzman$^{\ast\ast\ast}$ and Jorge Luis Acosta$^{\ast\ast\ast}$\bigskip}

\address{$^{*}$ Departamento de F\'isica Te\'orica, Instituto de Cibern\'etica, Matem\'atica y F\'isica (ICIMAF), La Habana, Cuba. \bigskip }
\address{$^{**}$  Facultad de Matem\'aticas y Computaci\'on, \\Universidad de La Habana, La Habana, Cuba. \bigskip}
\address{$^{***}$  Instituto de Tecnolog\'ias y Ciencias Aplicadas (InSTEC),\\ La Habana,Cuba.\bigskip }

\begin{abstract}
\noindent The possibility  that the spontaneous symmetry breaking in the Standard Model (SM) may be generated by the Top-Higgs Yukawa interaction (which determines the so called ``second minimum" in the SM) is examined. A former analysis is extended about a QCD action only including the Yukawa interaction of a single quark with a scalar field. We repeat the calculation of the two loop effective action of the model for the scalar field. A correction of the evaluation allowed choosing a strong coupling  $\alpha $($\mu,\Lambda_{QCD})=0.2254$ GeV at an  intermediate  scale $\mu=11.63$ GeV, in order to fix the minimum of the potential at a scalar field determining $175$ GeV for the quark
mass. A scalar field mass $m=44$ GeV is following,  which is of the order than the experimental Higgs mass. The effects of considering a running with momenta coupling are studied. For this,
the finite part of the two loop potential contribution determined by the strong coupling, was represented as a momentum integral. Next, substituting in this integral the experimental values of the running coupling, the potential curve became very similar to the one for
constant coupling. This happened after simply assuming that the low momentum
dependence of the coupling is "saturated" to a constant value being close to
its  lowest  experimental value.

\end{abstract}
\maketitle

\section{Introduction}

The so called "second minimum" of the Higgs field potential in the Standard
Model is the result of the Yukawa interaction of the Higgs field with the Top
quark. The presence of that minimum had been intrinsically related with the
same construction of the model along the years. Special procedures of fixing
the various parameters of the theory had to be designed in order to assure
that the minimum is separated from the usual Higgs extremum by a potential
barrier, being impossible to be tunneled by the standard physical vacuum
\cite{second-1,second-2,second-3}. In addition proposals had been advanced
that determine the Higgs mass from the condition for the two minima to
coincide in values of the potential \cite{second-3}. In Ref. \cite{cabo} a
simple massless QCD model including only one quark type (modeling the Top
quark) and a scalar field (modeling the Higgs field) with a Yukawa interaction
between them, was investigated. The aim of the study was to explore a
suspicion about that the so called the "second minimum" could in fact be the
responsible for the symmetry breaking in the SM. The idea was to evaluate the
two loop effective potential for the scalar field, which in the SM is the
responsible for the generation of the "second minimum" and to study the
possibility of choosing the renormalization conditions to fix the value of the
single fermion mass as equal to the top quark one $175$ GeV. An idea strongly
motivating this previous work, came after noting that this additional minimum
was identified only after the SM calculations arrived up to the two loop
order. Then, the question emerges about what could had been the result of an
attempt to construct the SM around this new radiative corrections determined
minimum, if it would had been known from the start in the SM construction. Up
to our knowledge, there had not been attempts to answer this question in the
past literature. The results in Ref. \cite{cabo} were inconclusive, in spite
of the fact the correct experimental values of the Higgs and the Top quark
masses were able to be fixed by choosing a definite value of the strong
coupling parameter. However, it happened, that the calculated value of this
parameter was a high one: $\alpha=\frac{g^{2}}{4\pi}$ close to $1,$ which
assuming the one loop formula for the relation between the coupling and the
scale corresponded to a low momentum scale $\mu=0.49$ GeV, being outside the
region of measured experimental values of the couplings.

In the present work we extend the study done in reference \cite{cabo}. The
discussion starts by considering a new evaluation of the two loop effective
potential for the mean value of the scalar field modeling the Higgs. The
discussion will proceed in two main directions. \noindent1) The first one is
to reconsider the two loop evaluation done in \cite{cabo} in order to search
for possible faults in those calculations, which could had altered the
obtained numerical values of the couplings and the scale required for fixing
the Top quark mass to its observed value.

\noindent2) In second place we will also consider to employ the running values
of the coupling with the momentum in the evaluation, in order to check if the
decreasing values of the coupling with momentum, also allows to justify the
fixing of the potential minimum to reproduce the Top quark mass, which was
attained at constant coupling.

In connection with the new evaluation of the potential, we present the results
of the calculation of the three relevant loop integrals determining the
effective potential for the scalar field. The revision allowed to detected a
numerical error which slightly affected the calculated coupling and scale
values for fixing the Top mass. The corrected results were employed to
calculate the new values of the coupling and the scale. The change resulted a
positive one: the new scale and coupling values (which were assumed to be
related by the one loop formula for the coupling) resulted in values being
larger for the scale: $\mu=11.63$ GeV with respect to the value $\mu=0.49$ GeV
evaluated in \cite{cabo}. For coupling values the new result was $\alpha
=\frac{g^{2}}{4\pi}=0.225445$, a smaller result than the high outcome of
nearly $\alpha\simeq1$\ following in the former work. Then, this first
conclusion support the suspected possibility that the spontaneous symmetry
breaking in the SM could be generated only by the Top quark-Higgs Yukawa interaction.

In order to consider the use of the running coupling with the momentum in
evaluating the potential, we firstly reformulated the finite integral defining
the quark-gluon effective potential contribution, which is directly determined
by the strong coupling (the quark loop with a contracted gluon propagator).
After substracting specially designed divergent parts of the relevant Feynman
integral, it was possible to identically transform its finite part in the
Minimal Substraction scheme in an integral over the momenta. This technical
result directly allowed to substitute the constant strong coupling by the
running with the momentum one in the integral.

The result showed that by simply assuming that the coupling dependence on the
momentum is "saturated" to a constant value for momentum smaller the smallest
of the measured momenta at which the running coupling is experimentally
measured, the calculated component of the effective potential becomes very
close numerically to the one evaluated at the initial constant coupling
$g(\mu,\Lambda_{QCD})$ at $\mu=11.63$ GeV. This result indicates that the
diminishing of the coupling with momentum does not alter the result for the
effective potential, which shows a minimum at a scalar field mean value
imposing a Top quark mass of $175$ GeV.

Another important result, is that the new formula for the effective potential
shows a second derivative at its minimum which predicts a scalar field mass of
nearly $m=44$ GeV. This result is smaller but yet close to the observed Higgs
mass of $126$ GeV. The new value corrects the one evaluated  in Ref.
\cite{cabo}. We estimate this conclusion as one interesting outcome of the
analysis. It means that once the Top quark mass is fixed, the spontaneous
symmetry breaking pattern associated to the Top-Higgs Yukawa interaction (that
is to the "second minimum") is able to determine a mass value of the scalar
field being close to the experimentally measured mass of the Higgs particle.
Therefore, it might be expected that the many new contributions to the
curvature of the Higgs potential that will exist in a more realistic SM type
of calculation make feasible to obtain the experimental value of $126$ GeV for
the Higgs mass. Therefore, the discussion in the work still sustain the
expectation about the possibility of describing the full SM after considering
an initial Lagrangian in which the classical Mexican hat potential may be
absent. The exploration of this possibility will considered elsewhere.

The plan of the work is as follows. In Section 2, the model and its Feynman
expansion are reviewed. Section 3 continues by presenting the new evaluation
of the effective potential for the mean scalar, and discussing the changes
with respect to the previous calculations in Ref. \cite{cabo}. Section 4
exposes the determination of the new values of the scale parameter $\mu=11.63$
GeV and its associated strong coupling value which allowed to fix the Top mass
as equal to the experimental value. Next, Section 5 describes the derivation
of the transformation of the effective potential contribution depending on the
strong coupling, in a momentum integral. This allows to substitute the
constant value of the strong coupling by the running with the momentum formula
in Section 6. Finally, the results are reviewed at the Summary.

\section{The model}

Let us now start by reviewing the main elements of the model discussed in
\cite{cabo}. The generating functional of the Feynamn expansion is based in an
action including a  singlet scalar field interacting with only one type of
quark. The functional was chosen in the form
\begin{equation}
Z[j,\eta,\overline{\eta},\xi,\overline{\xi},\rho]=\frac{1}{\mathcal{N}}%
\int\mathcal{D}[A,\overline{\Psi},\Psi,\overline{c},c,\phi]\exp[i\text{
}S[A,\overline{\Psi},\Psi,\overline{c},c,\phi]].\label{Z}%
\end{equation}
The action was taken in the form written below, in which in addition to the
usual massless QCD Lagrangian, there were only considered a Yukawa interaction
term  of a quark with a one component   scalar field and the corresponding
action term for the scalar. To simplify the discussion, the free action of the
scalar field was defined as a massless free term in the absence of
self-interaction. The action, after decomposed in its free and interaction
parts, is written below
\begin{align}
S &  =\int dx(\mathcal{L}_{0}\mathcal{+L}_{1}\mathcal{)},\\
\mathcal{L}_{0} &  =\mathcal{L}^{g}+\mathcal{L}^{gh}+\mathcal{L}%
^{q}+\mathcal{L}^{\phi},\\
\mathcal{L}^{g} &  =-\frac{1}{4}(\partial_{\mu}A_{\nu}^{a}-\partial_{\nu
}A_{\mu}^{a})(\partial^{\mu}A^{a,\nu}-\partial^{\nu}A^{a,\mu})-\frac
{1}{2\alpha}(\partial_{\mu}A^{\mu,a})(\partial^{\nu}A_{\nu}^{a}),\\
\mathcal{L}^{gh} &  =(\partial^{\mu}\chi^{\ast a})\partial_{\mu}\chi
^{a},\label{S}\\
\mathcal{L}^{q} &  =\overline{\Psi}\text{ }i\gamma^{\mu}\partial_{\mu}\text{
}\Psi,\\
\mathcal{L}^{\phi} &  =\frac{1}{2}\partial^{\mu}\phi\partial_{\mu}\phi,\\
\mathcal{L}_{1} &  =-\frac{g}{2}f^{abc}(\partial_{\mu}A_{\nu}^{a}%
-\partial_{\nu}A_{\mu}^{a})A^{b,\mu}A^{c,\nu}-g^{2}f^{abe}f^{cde}A_{\mu}%
^{a}A_{\nu}^{b}A^{c,\mu}A^{d,\nu}-\nonumber\\
&  -gf^{abc}(\partial^{\mu}\chi^{\ast a})\chi^{b}A_{\mu}^{c}+g\overline{\Psi
}T^{a}\gamma^{\mu}\Psi A_{\mu}^{a}+y\text{ }\overline{\Psi}\Psi\text{ }\phi.
\end{align}
\newline The dimensionless Yukawa coupling $y$ will be assumed to have a value
close to $y=1$ as it had been estimated in the literature \cite{castano}.
After constructing the Feynman expansion being associated to the above
generating function and classical action, the evaluation of the effective
potential as a function of an homogeneous scalar (Higgs resembling) field was
considered in reference \cite{cabo}, up to the two loop approximation. All the
notations for the quantum fields quantities, Minkowski metric, etc. used in
this work closely follow the ones employed in reference \cite{muta} .

\section{Two loops effective potential of the scalar field}

Let us consider again  the evaluation of all the contributions to the
effective potential $V(\phi)$ for the scalar field $\phi,$ up to the two loop
order. This is the quantity determining the spontaneous symmetry breaking
predictions of the considered model and checking its calculation is central
for to be sure about its physical predictions. We will see that numerical
errors slightly affected the results of the previous work. The corrections
will then allow to modify the results for the scale parameter $\mu$ and the
coupling $g(\mu,\Lambda_{QCD})$ values required for fixing the Top quark mass
value for the fermion in the model.

\subsection{ The one loop term\ }

\ The analytic expression for the one loop contribution shown in Fig. 1 which
was evaluated in \cite{cabo}, was given by the classical logarithm of the
fermion quark determinant as:
\begin{align}
\Gamma^{(1)}[\phi] &  =-V^{(D)}N\int\frac{dp^{D}}{i\text{ }(2\pi)^{D}%
}Log[Det\text{ }(G_{ii^{\prime}}^{(0)rr^{\prime}}(\phi,p))],\\
D &  =4-2\epsilon,\nonumber
\end{align}
\begin{figure}[h]
\includegraphics[width=7cm]{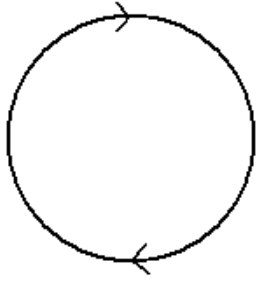}\caption{ The figure shows the quark one
loop correction. The result depends on the scalar \textquotedblright
mass\textquotedblright\ field $\phi$ through the quark free propagator which
is the usual free Green function of QCD, in which the mass is substituted by
$\phi$.}%
\end{figure}where $D$ is the space dimension of dimensional regularization and
the free fermion propagator was written as before in the conventions of Ref.
\ \cite{muta}, which, as mentioned before, will be used also throughout this
work. This propagator is defined as
\begin{align}
G_{ii^{\prime}}^{(0)rr^{\prime}}(\phi,p) &  =\delta^{ii^{\prime}}%
{\Large (}\frac{1}{-p_{\mu}\gamma^{\mu}+\phi}{\Large )}^{rr^{\prime}%
}\nonumber\\
&  =-\frac{\delta^{ii^{\prime}}}{p^{2}-\phi^{2}}{\Large (}p_{\mu}\gamma^{\mu
}+\phi{\Large )}^{rr^{\prime}}.\label{mqdef}%
\end{align}
As before, assuming the case of QCD with $SU(N)$ symmetry for $N=3$, and
evaluating the spinor and color traces, the one loop expression is simplified
to become
\[
\Gamma^{(1)}[\phi]=V^{(D)}\frac{N}{2}\int\frac{dp^{D}}{i\text{ }(2\pi)^{D}%
}Log[(\phi^{2}-p^{2})^{4}].
\]
Taking the derivative over $\phi^{2}$ of $\Gamma^{(1)}[\phi]$ allows to write
the easily integrable expression
\[
\frac{d}{d\text{ }\phi^{2}}\Gamma^{(1)}[\phi]=V^{(D)}2N\int\frac{dp^{D}%
}{\text{ }(2\pi)^{D}}\frac{1}{(p^{2}+\phi^{2})}.
\]
Making use of the identity
\begin{equation}
\int\frac{dp^{D}}{\text{ }(2\pi)^{D}}\frac{1}{(p^{2}+\lambda^{2})}%
=\frac{\Gamma(1-\frac{D}{2})}{(4\pi)^{\frac{D}{2}}}(\lambda^{2})^{\frac{D}%
{2}-1},\label{oneloop}%
\end{equation}
and integrating the result back over $\phi^{2}$, gives the dimensionally
regularized expression
\begin{align}
\Gamma^{(1)}[\phi] &  =V^{(D)}\frac{2N\Gamma(1-\frac{D}{2})}{(\frac{D}%
{2})(4\pi)^{\frac{D}{2}}}(\phi^{2})^{\frac{D}{2}-1}\nonumber\\
&  =V^{(D)}\frac{2N\text{ }\Gamma(\epsilon-1)}{(\frac{D}{2})(4\pi
)^{2-\epsilon}}(\phi^{2})^{2-\epsilon},
\end{align}
which coincides with the corresponding expression in \cite{cabo}. Let us
divide $\Gamma^{(1)}[\phi]$ by $\frac{V(D)}{\mu^{2\epsilon}}$, in order to
write the action density. The quantity $\mu$ in the denominator is the
dimensional regularization scale parameter, and the divisor $\mu^{2\epsilon},$
which tends to one on removing the regularization, is introduced in order
avoid results containing logarithms of quantities having dimension. Then, the
one loop Lagrangian density takes the form
\begin{equation}
\gamma^{(1)}[\phi]=\frac{\Gamma^{(1)}[\phi]}{\frac{V(D)}{\mu^{2\epsilon}}%
}=\frac{2N\text{ }\Gamma(\epsilon-1)}{(\frac{D}{2})(4\pi)^{2-\epsilon}}%
\phi^{4}(\frac{\phi}{\mu})^{-2\epsilon},
\end{equation}
also coinciding withe former result in \cite{cabo}. \ \ After deleting the
pole part of the above expression according to the Minimal Substraction rule,
and taking the limit $\epsilon\rightarrow0,$ gives the finite part of the one
loop action density as
\begin{equation}
\left[  \gamma^{(1)}[\phi]\right]  _{finite}^{\epsilon\rightarrow0}%
=\frac{3\phi^{4}}{32\pi^{2}}(-3+2\gamma-4\text{ }\log(2)-2\text{ }\log
(\pi)+2\text{ }\log(\frac{\phi^{2}}{\mu^{2}})),\label{e1}%
\end{equation}
where $\gamma=0.57721..$ is the Euler constant.\\ 
Finally, the one loop potential energy density is given by the negative of the
above quantity
\begin{align}
v^{(1)}[\phi] &  =-\frac{3\phi^{4}}{32\pi^{2}}(-3+2\gamma-4\text{ }%
\log(2)-2\text{ }\log(\pi)+2\text{ }\log(\frac{\phi^{2}}{\mu^{2}%
})),\nonumber\\
&  =-\frac{3\phi^{4}}{32\pi^{2}}(-3+2\gamma+2\text{ }\log(\frac{\phi^{2}}%
{4\pi\mu^{2}})).\label{e2}%
\end{align}

It can be noticed that one loop potential density is unbounded from below for
increasing values of the scalar field, which is its main property determining
the dynamical generation of the field $\phi$ \ in the \ model.

\subsection{ Quark-gluon two loop term}

\ Let us start now evaluating the two loop quark-gluon term which was
calculated in reference \cite{cabo} and  is illustrated in Fig. 2. Again,
after evaluating the color and spinor traces the analytic expression for this
contribution was obtained in a coinciding form as follows
\begin{equation}
\Gamma_{g}^{(2)}[\phi]=-V^{(D)}g^{2}(N^{2}-1)\int\frac{dp^{D}dq^{D}}%
{i^{2}\text{ }(2\pi)^{2D}}\frac{D\phi^{2}-(D-2)p.(p+q)}{q^{2}(p^{2}-\phi
^{2})((p+q)^{2}-\phi^{2})},\label{quarkgluon}%
\end{equation}
where similarly as before $g^{2}$ is the QCD coupling constant in the
dimensional regularization scheme, which introduces the scale parameter $\mu$
according to
\begin{equation}
g=g_{0}\text{ }\mu^{2-\frac{D}{2}}=g_{0\text{ }}\mu^{\epsilon}.
\end{equation}
\begin{figure}[h]
\includegraphics[width=7cm]{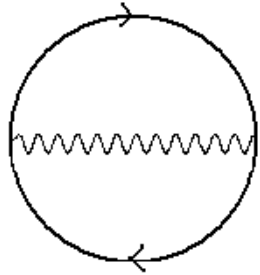} \caption{ The two loop contribution
determined by the strong interaction. As before, the $\phi$ dependence of the
result is introduced though the free quark propagator.}%
\label{fig2}%
\end{figure}After repeating the same steps followed in \cite{cabo}, that is:
symmetrizing the expression of $\Gamma_{g}^{(2)}[\phi]$ under the change of
sign in the momentum $q$, by means of the integration variable shift
$\ p\rightarrow p-\frac{q}{2}$ and the use of the identity
\begin{equation}
p^{2}=(p+\frac{q}{2})^{2}-\phi^{2}+\phi^{2}-\frac{q^{2}}{4}%
-q.p,\label{pcuadrado}%
\end{equation}
the quark-gluon term is written in the form
\begin{equation}
\Gamma_{g}^{(2)}[\phi]=\Gamma_{g}^{(2.1)}[\phi]+\Gamma_{g}^{(2,2)}[\phi],
\end{equation}
where $\Gamma_{g}^{(2,1)}[\phi]$ and $\Gamma_{g}^{(2,2)}[\phi]$ have the
formulae
\begin{align}
\Gamma_{g}^{(2,1)}[\phi] &  =-V^{(D)}2\phi^{2}g^{2}(N^{2}-1)\int\frac
{dk_{1}^{D}dk_{2}^{D}}{i^{2}\text{ }(2\pi)^{2D}}\frac{1}{k_{1}^{2}(k_{2}%
^{2}-\phi^{2})((k_{1}+k_{2})^{2}-\phi^{2})}\nonumber\\
&  =-V^{(D)}\frac{2\phi^{2}g^{2}(N^{2}-1)}{i^{2}\text{ }(2\pi)^{2D}}%
J_{111}(0,\phi,\phi),\\
\Gamma_{g}^{(2,2)}[\phi] &  =-V^{(D)}\frac{(D-2)g^{2}(N^{2}-1)}{2i^{2}%
(2\pi)^{2D}}{\LARGE (}\int dk_{1}^{D}\frac{1}{k_{1}^{2}-\phi^{2}}%
{\LARGE )}^{2},\label{squared}%
\end{align}
in which, the master two loop integral $J_{111}(0,\phi,\phi)$ was evaluated
making use of the results in Ref. \cite{fleischer}, and its explicit form for
the particular values of our arguments is:
\begin{align}
J_{111}(0,\phi,\phi) &  =\int dk_{1}^{D}dk_{2}^{D}\frac{1}{k_{1}^{2}(k_{2}%
^{2}-\phi^{2})((k_{1}+k_{2})^{2}-\phi^{2})}\nonumber\\
&  =-\frac{A(\epsilon)\pi^{4-2\epsilon}}{\epsilon^{2}}(\phi^{2})^{1-2\epsilon
},\\
A(\epsilon) &  =\frac{(\Gamma(1+\epsilon))^{2}}{(1-\epsilon)(1-2\epsilon)}.
\end{align}
Then, using the above definitions the following expression can be written for
$\Gamma_{g}^{(2,1)}$
\begin{equation}
\Gamma_{g}^{(2,1)}[\phi]=-\frac{V^{(D)}2g_{0}^{2}\mu^{2\epsilon}(N^{2}%
-1)}{\text{ }(2\pi)^{8-4\epsilon}}\frac{A(\epsilon)\pi^{4-2\epsilon}}%
{\epsilon^{2}}\phi^{4}(\phi^{2})^{-2\epsilon},
\end{equation}
and coincides with the former result.

However, for the case of $\Gamma_{g}^{(2,2)},$ we found that the result in
\cite{cabo} included an error after the square of the one loop integral in
equation (\ref{oneloop}) was substituted: a minus sign comming from the Wick
rotation of the squared momentum integral in (\ref{squared}) was ommitted.
\ Thus, the correct expression for this term should be
\begin{equation}
\Gamma_{g}^{(2,2)}[\phi]=-V^{(D)}\frac{g_{0}^{2}\mu^{2\epsilon}(N^{2}%
-1)2(1-\epsilon)}{\text{ }2(2\pi)^{8-4\epsilon}}\pi^{4-2\epsilon}(\Gamma
_{g}(\epsilon-1))^{2}\phi^{4}(\phi^{2})^{-2\epsilon}.
\end{equation}
Again dividing by $\frac{V(D)}{\mu^{2\epsilon}}$ to evaluate the action
densities gives
\begin{align}
\gamma_{g}^{(2,1)}[\phi] &  =\frac{\Gamma_{g}^{(2,1)}[\phi]}{\frac{V(D)}%
{\mu^{2\epsilon}}}\nonumber\\
&  =-\frac{2g_{0}^{2}(N^{2}-1)}{\text{ }(2\pi)^{2D}}\frac{A(\epsilon
)\pi^{4-2\epsilon}}{\epsilon^{2}}\phi^{4}(\frac{\phi^{2}}{\mu^{2}%
})^{-2\epsilon},
\end{align}
and for the corrected term
\begin{align}
\gamma_{g}^{(2,2)}[\phi] &  =\frac{\Gamma_{g}^{(2,2)}[\phi]}{\frac{V(D)}%
{\mu^{2\epsilon}}}\nonumber\\
&  =-V^{(D)}\frac{g_{0}^{2}\mu^{2\epsilon}(N^{2}-1)2(1-\epsilon)}{\text{
}2(2\pi)^{8-4\epsilon}}\pi^{4-2\epsilon}(\Gamma(\epsilon-1))^{2}\phi^{4}%
(\phi^{2})^{-2\epsilon}.
\end{align}

Therefore, repeating the process of substracting the divergent poles and
taking the limit $\epsilon\rightarrow0$  the total

quark-gluon two loop finite contribution to the action density takes the
expression
\begin{align}
\left[  \gamma_{g}^{(2)}[\phi]\right]  _{finite}^{\epsilon\rightarrow0} &
=-\frac{g_{0}^{2}}{64\pi^{4}}\phi^{4}{\LARGE (}30-28\gamma+12\gamma^{2}%
+\pi^{2}\text{ }+56\text{ }\log(2)\text{\ }-48\gamma\log(2)+\nonumber\\
&  48\text{ }\log(2)^{2}\text{\ }+28\text{ }\log(\pi)-24\gamma\log
(\pi)+48\text{ }\log(2)\text{ }\log(\pi)+12\text{ }\log(\pi)^{2}+\nonumber\\
&  (24\gamma-28-48\text{ }\log(2)-48\text{ }\log(\pi))\text{ \ }\log
(\frac{\phi^{2}}{\mu^{2}})+12\text{ }(\log(\frac{\phi^{2}}{\mu^{2}}%
))^{2}{\LARGE )}\\
&  =-v_{g}^{(2)}[\phi],\label{loopgluon}%
\end{align}
in which $v_{g}^{(2)}[\phi]$ defines the quark-gluon contribution to the
effective potential.

It should be remarked, that the leading logarithm squared term in the action
is negative, indicating that the contribution of the usual quark-gluon diagram
to the potential (equal to \ minus the action) up to the two loop
approximation remains being bounded from below as a function of $\phi,$
\ after the corrections are done.

The divergent contribution to the action follows in the form
\begin{equation}
\gamma_{g,\operatorname{div}}^{(2)}[\phi]=-\frac{3g^{2}\phi^{4}}{32\pi
^{4}\epsilon^{2}}+\frac{\text{ }g^{2}\phi^{4}}{32\text{ }\pi^{4}\epsilon
}{\Large (}-7+6\text{ }\gamma-6\log(4\pi)+12\log(\frac{\phi}{\mu}%
){\Large ),}\label{div}%
\end{equation}

which defines \ the Minimal Substraction  making finite \ the quark-gluon two
loops contribution.

\subsection{Scalar-quark two loop term}

Finally, let us repeat the evaluation \ of the two loop term being associated
to the quark-scalar loop illustrated in Fig. 3 . Due to the absence of spinor
and color structures in the vertices the analytic expression for this term is
again calculated to be \ \
\begin{equation}
\Gamma_{Y}^{(2)}[\phi]=V^{(D)}2N\int\frac{dp^{D}dq^{D}}{i^{2}\text{ }%
(2\pi)^{2D}}\frac{p^{2}-\frac{q^{2}}{4}+\phi^{2}}{q^{2}((p+\frac{q}{2}%
)^{2}-\phi^{2})((p-\frac{q}{2})^{2}-\phi^{2})},
\end{equation}
which in a close way as it was done for the quark-gluon term, was evaluated in
the form%
\begin{align}
\Gamma_{Y}^{(2)}[\phi] &  =V^{(D)}\frac{4N}{i^{2}\text{ }(2\pi)^{2D}}%
J_{111}(0,\phi,\phi)\nonumber\\
&  -V^{(D)}N{\LARGE (}\int\frac{dk_{1}^{D}}{i(2\pi)^{D}\text{ }}\frac{1}%
{k_{1}^{2}-\phi^{2}}{\LARGE )}^{2}\nonumber\\
&  =\frac{V^{(D)}4N}{\text{ }(2\pi)^{8-4\epsilon}}\frac{A(\epsilon
)\pi^{4-2\epsilon}}{\epsilon^{2}}\phi^{4}(\phi^{2})^{-2\epsilon}\nonumber\\
&  -V^{(D)}\frac{N}{\text{ }(2\pi)^{8-4\epsilon}}\pi^{4-2\epsilon}%
(\Gamma(\epsilon-1))^{2}\phi^{4}(\phi^{2})^{-2\epsilon}.
\end{align}
It can be noted that the imaginary number included in the squared momentum
integral, was now and before properly considered,  avoiding in this way the
error done in the former evaluation of the quark-gluon term. \begin{figure}[h]
\includegraphics[width=7cm]{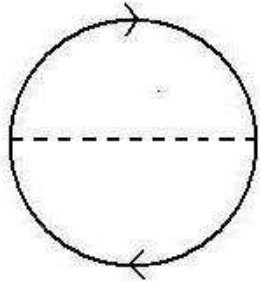}\caption{ The two loop contribution
determined by the scalar-quark interaction. The Yukawa coupling between the
quark and the scalar field was chosen as approximately equal to one.}%
\label{fig2}%
\end{figure}The division by the volume $\frac{V(D)}{\mu^{2\epsilon}}$ again
allows to write for the action density, the formula
\begin{align}
\gamma_{Y}^{(2)}[\phi] &  =\frac{4N}{\text{ }(2\pi)^{8-4\epsilon}}%
\frac{A(\epsilon)\pi^{4-2\epsilon}}{\epsilon^{2}}\phi^{4}(\frac{\phi^{2}}%
{\mu^{2}})^{-2\epsilon}\nonumber\\
&  -\frac{N}{\text{ }(2\pi)^{8-4\epsilon}}\pi^{4-2\epsilon}(\Gamma
(\epsilon-1))^{2}\phi^{4}(\frac{\phi^{2}}{\mu^{2}})^{-2\epsilon}.
\end{align}

Substracting the divergent pole part in $\epsilon$, passing to the limit
$\epsilon\rightarrow0$ gives for the potential density%
\begin{align}
v_{Y}^{(2)}[\phi] &  =-\left[  \gamma_{Y}^{(2)}[\phi]\right]  _{finite}%
^{\epsilon\rightarrow0}\nonumber\\
&  =-\frac{3}{512\pi^{4}}\phi^{4}{\LARGE (}50-40\gamma+12\gamma^{2}+\pi
^{2}\text{ }+96\text{ }\log(2)\text{\ }-64\gamma\log(2)+ \nonumber \\
&  64\text{ }\log(2)^{2}\text{\ }+48\text{ }\log(\pi)-32\gamma\log
(\pi)+64\text{ }\log(2)\text{ }\log(\pi)+16\text{ }\log(\pi)^{2}\nonumber\\
&  -8\text{ }\log(4\pi)+8\text{ }\gamma\text{ }\log(4\pi)-4\text{ }\log
(4\pi)^{2}+\nonumber\\
&  (24\gamma-40-64\text{ }\log(2)-32\text{ }\log(\pi)+8\log(4\pi))\text{
\ }\log(\frac{\phi^{2}}{\mu^{2}})+12\text{ }(\log(\frac{\phi^{2}}{\mu^{2}%
}))^{2}{\LARGE )}.\label{scalar}%
\end{align}
It can be noted that this contribution, being a two loop one, also includes a
squared logarithm term. However, its sign is contrary to the one appearing in
the quark gluon loop.

For the total two loop effective potential it follows
\begin{equation}
V[\phi,\mu]=v^{(1)}[\phi]+v_{g}^{(2)}[\phi]+v_{Y}^{(2)}[\phi].
\end{equation}

\section{ \ Fixing the potential minimum for m$_{Top}=175$ GeV}

Let us consider the sum of all the just evaluated contributions to the
potential energy density $V(\phi)$. Its expression is a combination of terms
of the form $\phi^{4}$, $\phi^{4}\log(\frac{\phi}{\mu})$ and $\phi^{4}%
(\log(\frac{\phi}{\mu}))^{2}$, with coefficients that only depend on the
strong coupling $g_{0}$ in the present first analysis. Then, in order to
approach the physical situation, we evaluated the potential $V(\phi)$ at the
values of $g_{0}$ satisfying the one loop formula for the running coupling
constant \cite{muta}
\begin{equation}
g_{0}(\mu,\Lambda_{QCD})=2\sqrt{\frac{2}{7}}\pi\sqrt{\frac{1}{\log(\frac{\mu
}{\Lambda_{QCD}})}}.
\end{equation}
The $\Lambda_{QCD}$ constant was chosen to be the estimate $\Lambda
_{QCD}=0.217\,\,GeV$. Here it should be remarked that for the determination of
the one loop coupling we have assumed the number of fermions as equal to six,
in place of to one, as it is proper for the model under discussion. This
criterion was adopted in order to assume the strong coupling as more
representative of the situation in the SM. In spite of this,  also had
evaluated the results for the case $N_{f}=1$ and the qualitative conclusions
of the work did not appreciably changed.

\begin{figure}[h]
\includegraphics[width=9cm]{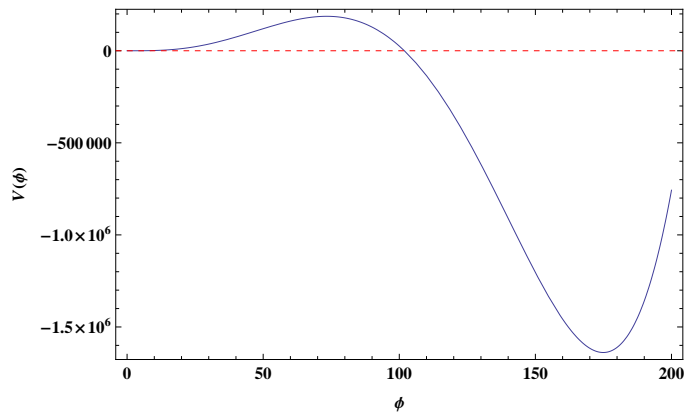}\caption{ The effective potential of the
mean field $\phi$ for the value of the scale $\mu$ determining  $\phi$ at the
potential minimum being equal to the top quark mass $m_{top}=175\,\,\,GeV$.
The second derivative at the minimum gives for the scalar field a low mass
$m_{\phi}=44\,\,GeV$, which is of the order the Higgs one $126$ $GeV$. In this
calculation the scale $\mu$ allowing the top mass fixation is within an
intermediate energy region: $\mu=11.63\,GeV$, which gives a coupling value
$\alpha=\frac{g_{0}^{2}}{4\pi}=0.225445$.}%
\label{fig4}%
\end{figure}

Next, we studied the potential curves in order to examine the behavior of
their minimum as functions of $\phi$, when the scale $\mu$ is changed. It
follows that when the strong coupling starts to increase as the scale diminish
down to one $GeV$, the value of $\phi$ at the minima, which determines the
quark mass also decreases. For the particular value of $\mu=11.63\,\,GeV$ ,
the potential curve is shown in Fig.\thinspace\ \ref{fig4}. The particular value of $\mu$ chosen, fixes the position of the
minimum at a field $\phi$ defining a quark mass of $~175$ $GeV$. The set of
parameters for this curve are
\begin{align}
\mu &  =11.63\,\,GeV,\\
g_{0} &  =1.68316\,\,\,\,(\alpha=\frac{(g_{0})^{2}}{4\pi}=0.225445),\\
\Lambda_{QCD} &  =0.217\,\,GeV.
\end{align}

\subsection{The mass of the scalar field}

Let us consider the mass of the scalar field to be defined in the present
calculation. For evaluating it, we write  the following approximate two loop
action for the scalar field linear propagation modes%
\begin{align}
\mathcal{L}^{\phi} &  =\frac{1}{2}\partial^{\mu}\phi\partial_{\mu}\phi
-\frac{1}{2}\phi V^{\prime\prime}[0]\phi,\\
V^{\prime\prime}[0] &  =\left.  \frac{\partial^{2}}{\partial\phi^{2}}%
V[\phi]\right\vert _{\phi=0}.
\end{align}

The Lagrange equation for the propagating scalar field waves $\phi=\exp(-i$
$p.x)$ then writes
\begin{equation}
(\partial^{\mu}\partial_{\mu}\phi+V^{\prime\prime}[0])\phi=(-p^{2}%
+V^{\prime\prime}[0])=0.
\end{equation}

Therefore the mass of the scalar field waves is given by%
\begin{equation}
m_{\phi}=\sqrt{V^{\prime\prime}[0]}.
\end{equation}
That is, the mass of the scalar field in a first approximation is defined by
the square root of the second derivative of the effective potential respect to
the mean field. Therefore, the second derivative of the potential curve in
Fig. 1, estimates for the mass of the scalar field $m_{\phi}=45$ $GeV$. This
value is smaller but of the order of the observed Higgs mass of $126$ GeV.
Then, after considering that by fixing the fermion mass to the top quark
experimental mass, had determined a mass  for the scalar field being close to
the Higgs's one, directly supports the possibility of generating the breaking
of symmetry in the SM through the Yukawa interaction between the Top quark and
the Higgs field. This possibility is also made plausible, by noticing that
upon considering a similar evaluation, but in a model showing the same field
content as the SM, there will exist a variety of additional particles. Some of
them also have similar masses  as the Top and Higgs (the W and Z bosons).
Thus, their contribution to the Higgs potential could correct the resulting
scalar field mass value to become close to the observed one.

As remarked before, the appeared single extremum of the potential is related
with the existence of so called "second minimum" of the Higgs potential in the
SM, laying at large values of the Higgs field. That minimum is recognized to
be produced precisely by the contributions of the top-quark Yukawa interaction
term, which is of the same form that  the one considered here
\cite{second-1,second-2,second-3}.

\section{Potential evaluation using the running coupling}

In this section, we will investigate the stability of the previous evaluation
of the effective potential for the scalar mean field by substituting the
constant value of the coupling chosen at a given scale by the running coupling
with momentum. The objective will be to check how robust can  be the fixing of
the Top quark mass, under the replacement of the constant value of the
coupling by a momentum dependent one. For this purpose the expression for the
finite part of the effective potential (obtained after employing the Minimal
Substractions scheme) will be represented as a momentum integral, in which the
replacement of the constant coupling can be afterwards implemented.

In fact this had been the most demanding technical part of the present work.
The difficulty was determined by the employed dimensional regularization
approach under the Minimal Substraction scheme. The obstacles were created by
the fact that normally, the full divergence structure of the evaluated
quantities near dimension equal to four, only appears after integrating over
the momenta. But, for approximately substituting the constant coupling by the
running one with momentum, it is required an expression for the finite part
being represented as a momentum integral. To derive this expression is the
main objective of most of the technical discussion to be presented below in
this section.

The plan of the section is as a follows. First we will present the formula for
the effective potential as represented by a momentum integral in terms of
Appell  series as functions of the momenta and $\epsilon=\frac{4-D}{2}.$ But,
since   the Euclidean space integral has a volume differential  of the form
$dV$=$dq$ $q^{3-2\epsilon}$ (determined by the $D$ dimensional integration
over the momenta), it was noted that for making finite the integral, it is
only needed to substract a specially designed  asymptotic form at large values
of $q$ of the factor $F(q)$ defining the full integrand as  $F(q)dV$.  The
complex form of these terms complicated the discussion, because the Appell
functions appearing do not show a pure Taylor expansion in powers of $\frac
{1}{q}.$ In fact, the expansion becomes a power series of $\frac{1}{q}$ with
factors which are powers of $q^{\epsilon}$. These factors,  although becoming
equals to 1 for $\epsilon\rightarrow0,$ contribute to the final result due to
the appearance of  divergent pole terms in $\epsilon.$

After substracting the appropriate divergent terms, a formula for a momentum
integral was obtained which became convergent at large momentum and showed a
single divergent term at small momentum as $\frac{1}{q}$. However, we noted
that after substituting few of these divergent factors as
\begin{equation}
\frac{1}{q}\rightarrow\frac{1}{\sqrt{q^{2}+\delta^{2}},}%
\end{equation}
the integral became again convergent at large momentum, but also at zero
momentum. This led to the introduction of a new parameter which afterwards
played a helpful role. At this point it was possible to take the limit
$\epsilon\rightarrow0$ in the integral due to its finite character. Then, we
passed to study the divergent contribution which was substracted to make the
integral finite at large as well as for zero momenta. Since the momentum
dependence of this substraction became simpler than the original one, it was
possible to exactly evaluate the momentum integral, which allowed to determine
its pole structure in $\epsilon$ as well as its finite part.

As it should result to be the case, the pole part of the divergent integral
exactly reproduced the Minimal Substraction required to make the two loop
quark-gluon term finite. As for its finite part, it resulted as a function of
the scale parameter $\mu,$ the mean field $\phi,$ the strong coupling
constant, but also of the new parameter $\delta$ introduced for making the
substracted integral convergent at small momentum. At this point came the
helpful character of the regularization parameter $\delta:$ we selected its
value as a function $\ $of $\mu,$ $\phi$ and $g$ for to impose that the finite
part of the divergent integral vanishes for all values of $\mu,$ $\phi$ \ and
$g.$ This fixation of $\delta$ became possible and a real and positive
solution exists for the variety of values of $\mu,$ $\phi$ and $g.$

Therefore, it followed that the defined finite integral over the momenta
exactly coincides with the effective potential when the strong coupling is
constant. Hence, the obtained formula for effective potential can be used to
explore the effects of considering that the strong coupling of the gluons with
the quarks runs with the exchanged momenta, within the contribution associated
to the quark self energy loop, contracted with the gluon propagator.

\subsection{The two loops quark-gluon contribution to the effective potential
as a momentum integral}

As it was mentioned in the past subsection, we want now to investigate the
effects that could have on the results to assume that the strong coupling
varies with the magnitude of the exchanged momentum $q.$ The evaluation of the
effective potential done in the past section seems amenable of being
influenced by assuming the coupling to run with the momentum.

We followed a specific path in order to derive a momentum integral for the
finite part of the quark-gluon contribution to the effective action. The
outcome, in one hand coincides with the result when the coupling is momentum
independent and then it was employed to investigate the effects of
substituting the strong coupling by a running with the momentum expression in
the following section.

We start by considering that the effective action is given by the contraction
of the gluon polarization tensor and free gluon propagator. Then, the
polarization tensor was expressed through the  formula derived in the page 374
of reference \cite{muta}%
\begin{equation}
\Pi_{\mu\nu}^{ab}(q)=-\frac{4g^{2}\delta^{ab}}{(4\pi)^{2-\epsilon}}%
\Gamma(\epsilon)(q^{2}g_{\mu\nu}-q_{\mu}q_{\nu})\int_{0}^{1}dx\text{
}x(1-x){\Large (}\phi^{2}-x(1-x)q^{2}{\Large )}^{-\epsilon}.
\end{equation}

After contracting the  tensor with the gluon propagator%
\begin{equation}
D_{\mu\nu}^{ab}(q)=\frac{\delta^{ab}}{q^{2}}{\large (}g_{\mu\nu}%
-(1-\alpha)\frac{q_{\mu}q_{\nu}}{q^{2}}{\large ),}%
\end{equation}
the quark-gluon contribution to the effective action for the scalar field
(minus the effective potential), after evaluating the integral over the
variable $x$ got the expression
\begin{align}
\Gamma_g^{(2)}(\epsilon,g,\phi,\mu) &  =\int_{0}^{\infty}dq\int_{0}^{1}%
dx\frac{2^{-3+4\epsilon}g^{2}q^{3-2\epsilon}\pi^{-4+2\epsilon}(-3+2\epsilon
)\Gamma(\epsilon)(1-x)x(1-q^{2}(-1+x)x)^{-\epsilon}\phi^{4}(\frac{\phi}{\mu
})^{-4\epsilon}}{\Gamma(2-\epsilon)},\nonumber\\
&  =\int_{0}^{\infty}dq\text{ }L_g^{(2)}(q,\epsilon,g,\phi,\mu).
\end{align}
where it was defined the momentum integrand $L_g^{(2)}(q,\epsilon,g,\phi,\mu)$ having
the explicit form%
\begin{align}
L_g^{(2)}(q,\epsilon,g,\phi,\mu) &  =\frac{1}{3\Gamma(2-\epsilon)}g^{2}q^{3-2\epsilon
}4^{\epsilon}(2\pi)^{-4+2\epsilon}(3-2\epsilon)\text{ }\phi^{4}\text{ }%
(\frac{\phi}{\mu})^{-4\epsilon}\text{ }\Gamma(\epsilon)\times\nonumber\\
&  (-3\text{AppelF}_{1}[2,\epsilon,\epsilon,3,\frac{(-q^{2}+\sqrt
{q^{2}(4+q^{2})})}{2},\frac{(-q^{2}-\sqrt{q^{2}(4+q^{2})})}{2}]+\nonumber\\
&  -2\text{AppelF}_{1}[3,\epsilon,\epsilon,4,\frac{(-q^{2}+\sqrt{q^{2}%
(4+q^{2})})}{2},\frac{(-q^{2}-\sqrt{q^{2}(4+q^{2})})}{2}],
\end{align}
in terms of the Appell functions \cite{bateman,schloser}
\begin{equation}
\text{AppelF}_{1}[a,b_{1},b_{2},c,x,y]=\sum_{m=0}^{\infty}\sum_{n=0}^{\infty
}\frac{(a)_{m+n}(b_{1})_{m}(b_{2})_{n}}{m!n!(c)_{m+n}}x^{m}y^{n},
\end{equation}
in which the Pochhammer symbols are defined as
\begin{equation}
(a)_{n}=\frac{\Gamma(a+n)}{\Gamma(a)}.
\end{equation}

The $L_g^{(2)}$ function is not convergent at large momentum, and the pole parts of
its momentum integral as functions of the $\epsilon$ parameter define the
Minimal Substraction required to make the result finite. However, as it was
mentioned before, by substracting the asymptotic behavior of the Appell
functions at large momentum the integral can be made finite. The resulting
integrand after this substraction can be written as
\begin{align}
L_{sub}(q,\epsilon,g,\phi,\mu,\delta)  &  =\frac{1}{3\Gamma(2-\epsilon)}%
g^{2}q^{3-2\epsilon}4^{\epsilon}(2\pi)^{-4+2\epsilon}(3-2\epsilon)\text{ }%
\phi^{4}\text{ }(\frac{\phi}{\mu})^{-4\epsilon}\text{ }\Gamma(\epsilon
)\times\nonumber\\
&  (-3\text{ }(\text{AppelF}_{1}[2,\epsilon,\epsilon,3,\frac{(-q^{2}%
+\sqrt{q^{2}(4+q^{2})})}{2},\frac{(-k^{2}-\sqrt{q^{2}(4+q^{2})})}%
{2}]\nonumber\\
&  -\text{Appel23Sub}[q,\epsilon,\delta])+\nonumber\\
&  2\text{ }(\text{AppelF}_{1}[3,\epsilon,\epsilon,4,\frac{(-q^{2}+\sqrt
{q^{2}(4+q^{2})})}{2},\frac{(-q^{2}-\sqrt{q^{2}(4+q^{2})})}{2}-\nonumber\\
&  \text{Appel34Sub}[q,\epsilon,\delta])] \label{Lsub}%
\end{align}
where the substractions done are defined by the large momentum asymptotic form
of the two entering Appell functions given by the fomulae
\begin{align}
\text{Appel23Sub}[k,\epsilon,\delta]  &  =\frac{2^{-1+2\epsilon}\pi^{\frac
{3}{2}}q^{-2\epsilon}\csc(\pi(1-\epsilon))}{\Gamma(\frac{3}{2}-\epsilon
)\Gamma(\epsilon)}+\nonumber\\
&  (\frac{1}{q^{2}})^{1-\epsilon}q^{-2\epsilon}{\Huge (}\frac{2^{1+2\epsilon
}\pi^{\frac{3}{2}}q^{-2\epsilon}\epsilon\csc(\pi(1-\epsilon))}{\Gamma(\frac
{3}{2}-\epsilon)\Gamma(\epsilon)}-\nonumber\\
&  \frac{2\pi\csc(\pi(1-\epsilon))}{\Gamma(2-\epsilon)\Gamma(\epsilon)}%
+\frac{4^{\epsilon}\pi^{\frac{3}{2}}q^{-2\epsilon}\epsilon\csc(\pi
(1-\epsilon))}{\Gamma(\frac{3}{2}-\epsilon)\Gamma(1+\epsilon)}{\Huge )}%
+\nonumber\\
&  (\frac{1}{q^{2}})^{-\epsilon}\frac{q^{-2\epsilon}}{(q^{2}+\delta^{2}%
)}{\Huge (}\frac{4^{1+\epsilon}\pi^{\frac{3}{2}}q^{-2\epsilon}\epsilon\csc
(\pi(1-\epsilon))}{\Gamma(\frac{3}{2}-\epsilon)\Gamma(\epsilon)}+\nonumber\\
&  \frac{4^{1+\epsilon}\pi^{\frac{3}{2}}q^{-2\epsilon}\epsilon^{2}\csc
(\pi(1-\epsilon))}{\Gamma(\frac{3}{2}-\epsilon)\Gamma(\epsilon)}+\frac
{8\pi\csc(\pi(1-\epsilon))}{\Gamma(2-\epsilon)\Gamma(\epsilon)}-\nonumber\\
&  \frac{12\pi\csc(\pi(1-\epsilon))}{\Gamma(3-\epsilon)\Gamma(\epsilon)}%
+\frac{8\pi\epsilon\csc(\pi(1-\epsilon))}{\Gamma(3-\epsilon)\Gamma(\epsilon
)}-\nonumber\\
&  \frac{4^{1+\epsilon}\pi^{\frac{3}{2}}q^{-2\epsilon}\epsilon\csc
(\pi(1-\epsilon))}{\Gamma(\frac{3}{2}-\epsilon)\Gamma(1+\epsilon)}%
-\frac{4^{1+\epsilon}\pi^{\frac{3}{2}}q^{-2\epsilon}\epsilon^{2}\csc
(\pi(1-\epsilon))}{\Gamma(\frac{3}{2}-\epsilon)\Gamma(1+\epsilon)}+\nonumber\\
&  \frac{3\times4^{\epsilon}\pi^{\frac{3}{2}}q^{-2\epsilon}\epsilon\csc
(\pi(1-\epsilon))}{\Gamma(\frac{3}{2}-\epsilon)\Gamma(2+\epsilon)}%
+\frac{3\times4^{\epsilon}\pi^{\frac{3}{2}}q^{-2\epsilon}\epsilon^{2}\csc
(\pi(1-\epsilon))}{\Gamma(\frac{3}{2}-\epsilon)\Gamma(2+\epsilon)}{\Huge )},
\end{align}
and%
\begin{align}
\text{Appel34Sub}[q,\epsilon,\delta]  &  =\frac{3}{2q^{2}(1-\epsilon)}%
+\frac{3}{4}\text{Appel23Sub}[q,\epsilon,\delta]-\nonumber\\
&  \frac{3\times2^{-4+2\epsilon}\pi^{\frac{3}{2}}q^{-2\epsilon}\epsilon
^{2}\csc(\pi(2-\epsilon))}{(-1+\epsilon)\Gamma(\frac{5}{2}-\epsilon
)\Gamma(-1+\epsilon)}+2^{-2\epsilon}(\frac{1}{q^{2}})^{1-\epsilon}%
{\Huge (}\frac{3\times2^{-2+4\epsilon}\pi^{\frac{3}{2}}q^{-2\epsilon}\csc
(\pi(2-\epsilon))}{\Gamma(\frac{5}{2}-\epsilon)\Gamma(-1+\epsilon
)}-\nonumber\\
&  \frac{3\times2^{-3+4\epsilon}\pi^{\frac{3}{2}}q^{-4\epsilon}\csc
(\pi(2-\epsilon))}{\Gamma(\frac{5}{2}-\epsilon)\Gamma(\epsilon)}%
{\Huge )}+\frac{2^{-2\epsilon}q^{2\epsilon}}{(q^{2}+\delta^{2})}%
{\Huge (}-\frac{3\times2^{-1+4\epsilon}\pi^{\frac{3}{2}}q^{-4\epsilon}%
\epsilon\csc(\pi(2-\epsilon))}{\Gamma(\frac{5}{2}-\epsilon)\Gamma
(-1+\epsilon)}+\nonumber\\
&  \frac{3\times2^{2\epsilon}\text{ }\pi\text{ }q^{-2\epsilon}\csc
(\pi(2-\epsilon))}{(-1+\epsilon)\Gamma(3-\epsilon)\Gamma(-1+\epsilon)}%
+\frac{3\times2^{-1+4\epsilon}\text{ }\pi^{\frac{3}{2}}\text{ }q^{-4\epsilon
}\epsilon\csc(\pi(2-\epsilon))}{\Gamma(\frac{5}{2}-\epsilon)\Gamma(\epsilon
)}-\nonumber\\
&  +\frac{9\times2^{-3+4\epsilon}\pi^{\frac{3}{2}}q^{-4\epsilon}\epsilon
\csc(\pi(2-\epsilon))}{\Gamma(\frac{5}{2}-\epsilon)\Gamma(1+\epsilon
)}{\Huge )}.
\end{align}

In the above two expressions, it should be noted that the before defined
quantity $\delta$ is appearing in the denominators of the form $(q^{2}%
+\delta^{2})$. They, appeared after making the substitution $q\rightarrow
\frac{1}{\sqrt{q^{2}+\delta^{2}}}$ in a $q^{4}$ denominator of the only term
diverging as $\frac{1}{q}$ at zero momentum. This procedure eliminates the
mentioned zero omentum divergence, furnishes a simple momentum dependence of
the result and makes the result a function of the quantity $\delta$. The
substitution, on another hand does not disturbed the large momentum
convergence of the considered integral. Therefore, the substracted divergent
expression has the form
\begin{align}
L_{count}(q,\epsilon,g,\phi,\mu,\delta)  &  =\frac{1}{3\Gamma(2-\epsilon
)}g^{2}q^{3-2\epsilon}4^{\epsilon}(2\pi)^{-4+2\epsilon}(3-2\epsilon
)\Gamma(\epsilon)\text{ }\phi^{4}(\frac{\phi}{\mu})^{-4\epsilon}%
\times\nonumber\\
&  (-3\text{ Appel23Sub}(q,\epsilon,\delta)+2\text{Appel34Sub}(q,\epsilon
,\delta)).
\end{align}
Now, as remarked above, the relative simplicity of the obtained momentum
dependence of the substracted term, allows to exactly perform the momentum
integrals to obtain the result
\begin{align}
S_{count}(\epsilon,g,\phi,\mu,\delta)  &  =\int_{0}^{\infty}dq\text{
}L_{count}(q,\epsilon,g,\phi,\mu,\delta)\nonumber\\
&  ={\Huge (}4^{-5+3\epsilon}g^{2}\pi^{-\frac{3}{2}+2\epsilon}(\frac{1}%
{\delta^{2}})^{\epsilon}(-3+2\epsilon)\text{ }\phi^{4}\text{ }(\frac{\phi}%
{\mu})^{-4\epsilon}\csc(\pi\epsilon)^{2}\times\nonumber\\
&  {\Huge (}-4^{2+\epsilon}\pi^{\frac{3}{2}}(\frac{1}{\delta^{2}})^{\epsilon
}(-2+\epsilon)(-1+\epsilon)^{2}\epsilon(-1+2\epsilon)\csc(2\pi\epsilon
)\Gamma(2-2\epsilon)\Gamma(1+\epsilon)-\nonumber\\
&  2^{1+2\epsilon}\pi^{\frac{3}{2}}(\frac{1}{\delta^{2}})^{\epsilon
}(-2+\epsilon)(-1+2\epsilon)(-3+4\epsilon)\csc(2\pi\epsilon)\Gamma
(3-2\epsilon)\Gamma(1+\epsilon)-\nonumber\\
&  4^{\epsilon}\pi^{\frac{3}{2}}(\frac{1}{\delta^{2}})^{\epsilon}%
(-1+2\epsilon)(1+4\epsilon)\csc(2\pi\epsilon)\Gamma(5-2\epsilon)\Gamma
(1+\epsilon)-\nonumber\\
&  64\text{ }(-2+\epsilon)(-1+\epsilon)^{2}\Gamma(2-2\epsilon)\Gamma(\frac
{5}{2}-\epsilon)\Gamma(-1+\epsilon)\Gamma(1+\epsilon)+\nonumber\\
&  64\text{ }(-1+\epsilon)^{3}\Gamma(2-2\epsilon)\Gamma(\frac{5}{2}%
-\epsilon)\Gamma(-1+\epsilon)\Gamma(1+\epsilon)+\nonumber\\
&  4^{1+\epsilon}\pi^{\frac{3}{2}}(\frac{1}{\delta^{2}})^{\epsilon}%
\epsilon(-1+2\epsilon)\csc(2\pi\epsilon)\Gamma(5-2\epsilon)\Gamma
(2+\epsilon){\Huge ))/}\nonumber\\
&  {\Huge (}\Gamma(\frac{3}{2}-\epsilon)\Gamma(2-2\epsilon)^{2}\Gamma(\frac
{5}{2}-\epsilon)\Gamma(3-\epsilon)\Gamma(-1+\epsilon)\Gamma(1+\epsilon
){\Huge ).}%
\end{align}

But, expanding this relation in series of the $\epsilon$ parameter leads to
the result
\begin{align}
S_{count}(\epsilon,g,\phi,\mu,\delta) &  =%
{\displaystyle\sum\limits_{n=-\infty}^{\infty}}
S_{count}^{(n)}(g,\phi,\mu,\delta)\text{ }\epsilon^{n}\nonumber\\
&  =-\frac{3g^{2}\phi^{4}}{32\pi^{4}\epsilon^{2}}+\frac{\text{ }g^{2}\phi^{4}%
}{32\text{ }\pi^{4}\epsilon}{\Large (}-7+6\text{ }\gamma-6\log(4\pi
)+12\log(\frac{\phi}{\mu}){\Large )}+\nonumber\\
&  \frac{\text{ }g^{2}\phi^{4}}{64\text{ }\pi^{4}}\times{\Large (}-21+\pi
^{2}+10\gamma-48\log(2)^{2}-28\log(4\pi)+\nonumber\\
&  6{\Large (}\gamma^{2}+\gamma(3-2\gamma)-\log(\pi)\log(256\pi^{2}%
)+\gamma(-\gamma+\log(256\pi^{4})){\Large )}-\nonumber\\
&  18\log(\frac{1}{\delta^{2}})+6\log(\frac{1}{\delta^{2}})^{2}+8{\Large (}%
7-6\gamma+6\log(4\pi)-6\log(\frac{\phi}{\mu}){\Large )}\log(\frac{\phi}{\mu
}){\Large )}\nonumber\\
&  +O^{(1)}(\epsilon){\Large ,}%
\end{align}
where $O^{(1)}(\epsilon)$ is a function vanishing when $\epsilon\rightarrow0.$

In the above formula, it should be remarked that the pole part, which defines
the divergent contribution, exactly coincides with the Minimal Substraction
term (\ref{div}) required to make finite the quark-gluon cntribution to the
effective action. Now, it can noticed that the expression for the integral
$S_{count}(\epsilon,g,\phi,\mu,\delta)$ (which was substracted from the
momentum integral of the term $L(q,\epsilon,g,\phi,\mu)$ to obtain a finite
remaining integral) has a finite part when $\epsilon\rightarrow0$. But, this
finite part is depending on the regularization parameter $\delta$ which was
used to make convergent the momentum integral around the zero momentum. This
circumstance opens the interesting possibility of choosing this value of
$\delta$ precisely to force the finite part to vanish for all the value of the
scale $\mu$ and the mean field. Then, imposing this condition for determining
$\delta,$ we may write
\begin{align}
&  \frac{\text{ }g^{2}\phi^{4}}{64\text{ }\pi^{4}}\times{\Large (}-21+\pi
^{2}+10\gamma-48\log(2)^{2}-28\log(4\pi)+\nonumber\\
&  6{\Large (}\gamma^{2}+\gamma(3-2\gamma)-\log(\pi)\log(256\pi^{2}%
)+\gamma(-\gamma+\log(256\pi^{4})){\Large )}-\nonumber\\
&  18\log(\frac{1}{\delta^{2}})+6\log(\frac{1}{\delta^{2}})^{2}+8{\Large (}%
7-6\gamma+6\log(4\pi)-6\log(\frac{\phi}{\mu}){\Large )}\log(\frac{\phi}{\mu
}){\Large )}=0{\Large .}%
\end{align}

One helpful property of this equation is the fact that it does not involve the
values of the strong coupling. This means that $\delta$ is only a function of
the scale $\mu$ and the mean field $\phi.$ The equation for $\delta$ has a
real and positive solution defined for all the values of the ratio $\frac
{\phi}{\mu}$ which can be expressed as follows
\begin{align}
\delta(\phi,\mu) &  =\frac{1}{\sqrt{\exp(\text{ }f_{1}\text{ }(36+4\text{
}\mathcal{F}\text{ }\mathcal{(\phi},\mathcal{\mu)}))}},\\
\mathcal{F}\text{ }\mathcal{(\phi},\mathcal{\mu)} &  \mathcal{=}\sqrt
{f_{2}+f_{3}\text{ }\log(\frac{\phi}{\mu})+f_{4}\text{ }\log(\frac{\phi}{\mu
})^{2}},\\
f_{1} &  =0.041666,\\
f_{2} &  =750.872709,\\
f_{3} &  =-898.696871,\\
f_{4} &  =+288.00.
\end{align}

\begin{figure}[h]
\includegraphics[width=9cm]{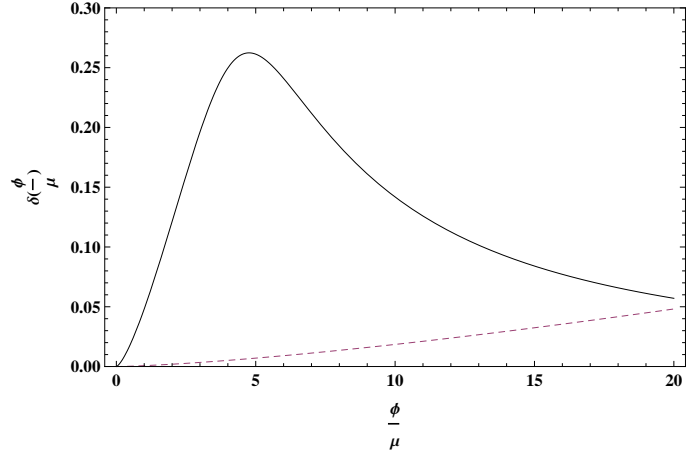}\caption{The figure shows the plot of
the real and positive solution for $\delta(\phi,\mu)$ as a function of the
ratio $\frac{\phi}{\mu}$. \ The dashed plot is the same curve in which the
horizontal axes is magnified in a factor of 20, to show the behavior near the
origin. }%
\label{delta}%
\end{figure}

The dependence of $\delta$ of  the ratio $\frac{\phi}{\mu}$ is depicted in
figure \ref{delta}.

Further, the function $\delta(\phi,\mu)$ was substituted in the integrand
$L_{sub}$ defined in equation (\ref{Lsub}) which upon integration furnishes
the finite integral. The result for the integrand of the finite integral can
then be written in the form,
\begin{align}
L_{MS}\text{ }(q,g,\phi,\mu) &  =\left.  L_{sub}(q,\epsilon,g,\phi,\mu
,\delta)\right\vert _{\delta\rightarrow\delta(\phi,\mu),\text{ }%
\epsilon\rightarrow0}\nonumber\\
&  =\frac{g^{2}\phi^{4}}{32\pi^{4}q}{\Huge (}12.9266-12\text{ }q^{2}%
+2.1789\text{ }q^{4}-12\text{ }\gamma+2\text{ }\gamma\text{ }q^{4}+\\
&  +\frac{0.643577-0.597445\gamma-0.597445\log(\frac{1}{q^{2}})}%
{{\Huge (}0.22313+\exp{\Large (}0.166667\sqrt{750.873+\log(\frac{\phi}{\mu
}){\large (}-898.697+288\log(\frac{\phi}{\mu}){\large )}}{\Large )}\text{
}q^{2}{\Huge )}^{2}}-\nonumber\\
&  12\log(\frac{1}{q^{2}})+6\text{ }q^{2}\log(\frac{1}{q^{2}})+\nonumber\\
&  \frac{-5.76862+5.35512\text{ }\gamma+5.35512\log(\frac{1}{q^{2}}%
)}{0.22313+\exp{\Large (}0.166667\sqrt{750.873+\log(\frac{\phi}{\mu
}){\large (}-898.697+288\log(\frac{\phi}{\mu}){\large )}}{\Large )}\text{
}q^{2}}+\nonumber\\
&  12\text{ }q^{2}\log(q)-4\text{ }q^{4}\log(q)+\nonumber\\
&  q^{4}\text{ }{\huge (}-6\text{ AppellF}_{1}^{(0,0,1,0,0,0)}{\Large (}%
2,0,0,3,\frac{1}{2}{\large (}-q^{2}+\sqrt{q^{2}(4+q^{2})}{\large )},\frac
{1}{2}{\large (}-q^{2}-\sqrt{q^{2}(4+q^{2})}{\large )}{\Large )}+\nonumber\\
&  4\text{ AppellF}_{1}^{(0,0,1,0,0,0)}{\Large (}3,0,0,4,\frac{1}{2}%
{\large (}-q^{2}+\sqrt{q^{2}(4+q^{2})}{\large )},\frac{1}{2}{\large (}%
-q^{2}-\sqrt{q^{2}(4+q^{2})}{\large )}\text{ }{\Large )}-\nonumber\\
&  -6\text{ AppellF}_{1}^{(0,1,0,0,0,0)}{\Large (}2,0,0,3,\frac{1}%
{2}{\large (}-q^{2}+\sqrt{q^{2}(4+q^{2})}{\large )},\frac{1}{2}{\large (}%
-q^{2}-\sqrt{q^{2}(4+q^{2})}{\large )}{\Large )}+\nonumber\\
&  4\text{ AppellF}_{1}^{(0,,1,0,0,0,0)}{\Large (}3,0,0,4,\frac{1}%
{2}{\large (}-q^{2}+\sqrt{q^{2}(4+q^{2})}{\large )},\frac{1}{2}{\large (}%
-q^{2}-\sqrt{q^{2}(4+q^{2})}{\large )}\text{ }{\Large )}\text{ }%
{\huge )}\text{ }{\Huge )},
\end{align}
\noindent where a superindex of the form $(n_{1},n_{2},n_{3},n_{4},n_{5}%
,n_{6})$ in the appearing Appell functions represents the numbers of $n_{i}$,
$i=1,...,6$ of the derivatives over the corresponding six arguments of the
Appell functions. It should be noticed that in writing this expression, the
limit $\epsilon\rightarrow0$ was also chosen, as allowed by the finiteness of
the integral. Therefore, we have arrived to an expression of the finite part
of the effective action in the form
\begin{equation}
\mathcal{L}(g,\phi,\mu)=\int dq\text{ }L_{MS}\text{ }(q,g,\phi,\mu
).\label{formula}%
\end{equation}

\begin{figure}[h]
\includegraphics[width=9cm]{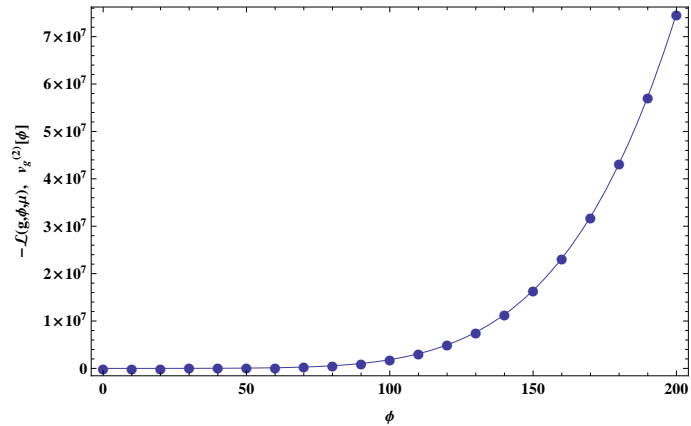}\caption{ The plots illustrate the
coincidence of the calculated  momentum integral representation for the
quark-gluon two loop contribution to the effective action with the result of
the direct evaluation of the same quantity. The solid curve indicates the
values of the momentum integral representation in (\ref{formula}) and the
dotted one the evaluation of the two loop quark gluon contribution in formula
(\ref{loopgluon}). }%
\label{comparison}%
\end{figure}

The figure \ref{comparison} shows the evaluations of the quark-gluon
contribution of the effective potential (minus the action) through both
formulae (\ref{loopgluon}) and (\ref{formula}). The coincidence of the two
plots checks the equivalence between the two expressions. The plots are done
for values of the scale parameter $\mu=11.63$ GeV and of the coupling
$g=g_{o}(11.63,\Lambda)$. The solid curve shows the values of $-\mathcal{L}%
(g,\phi,\mu)$ and the dotted one the values of $v_{g}^{(2)}[\phi]$ in
(\ref{loopgluon}) as functions of the mean field. 

In expression (\ref{formula}),  it is possible now  to replace the up to now
constant value of the strong coupling by its running expression with momentum.
The following section will discuss some properties of this substitution.

\section{The use of the running coupling with momentum}

In this section, the discussion starts by defining the running coupling to be
considered. Since it is known that the values of couplings are not well
defined at low momenta of the order of $1$ GeV, we defined values of the
running coupling "saturated " at low momentum. That is, in a neighborhood of
zero momentum they were assumed to be constant. Then, we explored the values
of the scale $\mu$ for which the substitution of the running coupling in the
obtained finite integral,  approximately coincide with the value attained at
constant value of the coupling for the same scale. Interestingly, it followed
that for the determined before value for the scale $\mu=11.63$ GeV, the effect
of the diminishing of the coupling with the momentum mainly does not affect
the obtained spontaneous symmetry breaking pattern.

For the purpose of substituting the constant coupling by its running
counterpart we analyzed two variants of couplings. The first of them was the
expression for the one loop renormalization coupling as a function of
momentum
\begin{align}
g_{o}(q) &  =%
\genfrac{\{}{.}{0pt}{}{\sqrt{\frac{1}{b_{o}\log(\frac{q^{2}}{\Lambda^{2}})}%
}\text{ \ , \ }\frac{q^{2}}{\Lambda^{2}}>\exp(\frac{16\pi^{2}}{7\text{
}g_{sat}^{2}})\text{ }}{\text{ \ \ \ \ \ \ \ }g_{sat}\text{ \ , \ \ \ \ \ }%
\frac{q^{2}}{\Lambda^{2}}<\exp(\frac{16\pi^{2}}{7\text{ }g_{sat}^{2}})\text{
}}%
,\\
g_{sat} &  =2.06702.
\end{align}
In this expression the couplings for momenta smaller than the value  at which
they become equal to the highest measured coupling $g_{sat}=2.06702$, are
assumed to remain constant, and equal to their "saturation" values
\cite{ParticleData}. Another form of the analyzed running coupling was given
by an interpolation of the set of experimental values reported in reference
\cite{ParticleData}. The expression describing the data was obtained in the
form%
\begin{equation}
g_{\exp}(q)=%
\genfrac{\{}{.}{0pt}{}{\left(  \log(\frac{23.4193}{32361.1672}q^{2})\right)
^{-1}\text{\ , \ \ }{\small q>1.6042285}}{\text{ \ \ \ \ \ \ \ \ \ }%
g_{sat},\text{ \ \ \ \ \ \ \ \ \ \ \ \ \ \ \ \ \ \ \ }{\small q<1.6042285}}%
,
\end{equation}
where $g_{sat}$ is the same saturation value defined before, that is, the
maximal value of the experimentally measured couplings given in reference
\cite{ParticleData}. \begin{figure}[h]
\includegraphics[width=9cm]{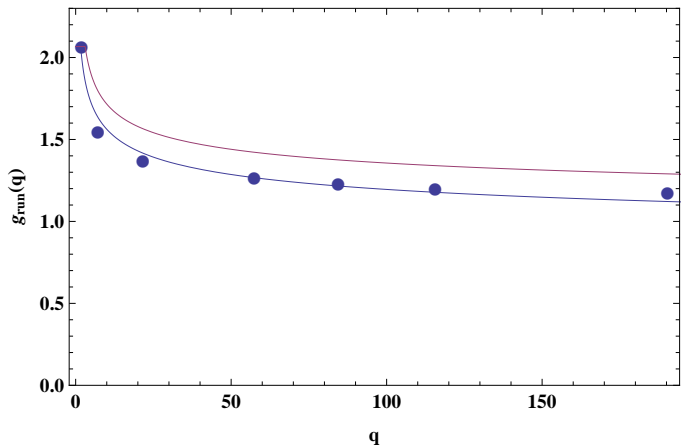} \caption{ The figure shows the
values of the fitting formula for a number of experimentally measured values
of the strong coupling according to reference \cite{ParticleData}. This curve
is the lower one and the fitted experimental points are indicated by the dots.
The higher plot shows the values of the one loop renormalization group running
coupling. }%
\label{couplings}%
\end{figure}Both coupling's momentum behavior are plotted in figure
\ref{couplings}. As it can noticed the values of the observations are
systematically smaller than the one loop determined values. Therefore, in what
follows we decided to employ the fitting curve of the experimental values for
substituting the constant coupling in formula (\ref{formula}).

\subsection{The quark-gluon effective potential evaluation using the running
coupling}

It is possible now to substitute the expression for the experimental value of
the running coupling $g_{\exp}(q)$ in the (\ref{formula}) to define the
quark-gluon contribution to the effective potential as evaluated at the
running coupling  values, in the form
\begin{equation}
V_{run}(\phi,\mu)=\int dq\text{ }\left.  L_{MS}\text{ }(q,g,\phi
,\mu)\right\vert _{g\rightarrow g_{\exp}(q)}.
\end{equation}

The resulting formula for this contribution to the effective potential takes
the form
\begin{align}
V_{run}(\phi,\mu)  &  =\int_{0}^{\infty}dq\text{ }\frac{g_{\exp}(q)^{2}%
\phi^{4}}{32\pi^{4}q}{\Huge (}12.9266-12\text{ }q^{2}+2.1789\text{ }%
q^{4}-12\text{ }\gamma+2\text{ }\gamma\text{ }q^{4}+\\
&  +\frac{0.643577-0.597445\gamma-0.597445\log(\frac{1}{q^{2}})}%
{{\Huge (}0.22313+\exp{\Large (}0.166667\sqrt{750.873+\log(\frac{\phi}{\mu
}){\large (}-898.697+288\log(\frac{\phi}{\mu}){\large )}}{\Large )}\text{
}q^{2}{\Huge )}^{2}}-\nonumber\\
&  12\log(\frac{1}{q^{2}})+6\text{ }q^{2}\log(\frac{1}{q^{2}})+\nonumber\\
&  \frac{-5.76862+5.35512\text{ }\gamma+5.35512\log(\frac{1}{q^{2}}%
)}{0.22313+\exp{\Large (}0.166667\sqrt{750.873+\log(\frac{\phi}{\mu
}){\large (}-898.697+288\log(\frac{\phi}{\mu}){\large )}}{\Large )}\text{
}q^{2}}+\nonumber\\
&  12\text{ }q^{2}\log(q)-4\text{ }q^{4}\log(q)+\nonumber\\
&  q^{4}\text{ }{\huge (}-6\text{ AppellF}_{1}^{(0,0,1,0,0,0)}{\Large (}%
2,0,0,3,\frac{1}{2}{\large (}-q^{2}+\sqrt{q^{2}(4+q^{2})}{\large )},\frac
{1}{2}{\large (}-q^{2}-\sqrt{q^{2}(4+q^{2})}{\large )}{\Large )}+\nonumber\\
&  4\text{ AppellF}_{1}^{(0,0,1,0,0,0)}{\Large (}3,0,0,4,\frac{1}{2}%
{\large (}-q^{2}+\sqrt{q^{2}(4+q^{2})}{\large )},\frac{1}{2}{\large (}%
-q^{2}-\sqrt{q^{2}(4+q^{2})}{\large )}\text{ }{\Large )}-\nonumber\\
&  -6\text{ AppellF}_{1}^{(0,1,0,0,0,0)}{\Large (}2,0,0,3,\frac{1}%
{2}{\large (}-q^{2}+\sqrt{q^{2}(4+q^{2})}{\large )},\frac{1}{2}{\large (}%
-q^{2}-\sqrt{q^{2}(4+q^{2})}{\large )}{\Large )}+\nonumber\\
&  4\text{ AppellF}_{1}^{(0,,1,0,0,0,0)}{\Large (}3,0,0,4,\frac{1}%
{2}{\large (}-q^{2}+\sqrt{q^{2}(4+q^{2})}{\large )},\frac{1}{2}{\large (}%
-q^{2}-\sqrt{q^{2}(4+q^{2})}{\large )}\text{ }{\Large )}\text{ }%
{\huge )}\text{ }{\Huge )}.
\end{align}

\begin{figure}[h]
\includegraphics[width=9cm]{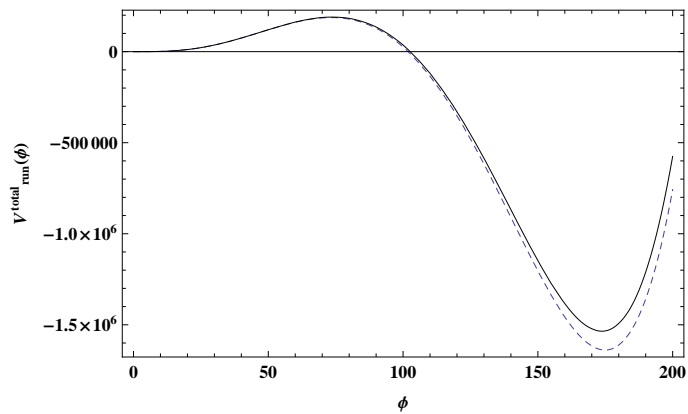}\caption{ The figure shows the plot of
the values of the total two-loop effective potential for the scalar field
(solid curve) after the running coupling is employed for calculating the
quark-gluon contribution. The dashed curve shows the similar potential as
evaluated for the constant values of the strong coupling. Note that the
results indicate that the momentum dependence of the coupling does not
appreciable disturb the arising spontaneous symmetry breaking pattern. }%
\label{Vrun}%
\end{figure}

Next, it possible to evaluate the effects of the running on the calculation of
the effective potential. Before, we have been able to fix the Top quark mass
by fixing the minimum of the potential (after calculated at constant strong
coupling) at the scale parameter value $\mu=11.63$ GeV. We then firstly
calculated $V_{run}(\phi,\mu)$ at this scale. The result of the evaluation of
the total effective potential
\begin{equation}
V_{run}^{total}(\phi,\mu)=v^{(1)}[\phi,\mu]+v_{Y}^{(2)}[\phi,\mu]+V_{run}%
(\phi,\mu),
\end{equation}
as a function of the mean scalar field in which the quark-gluon term is
calculated using the above formula for $V_{run}(\phi,\mu)$ is shown in figure
\ref{Vrun}. Note that the one loop and scalar two loop contributions are
defined by the same formulae (\ref{e2}) and (\ref{scalar}) which were used in
calculating the potential for constant coupling, since they do no depend on
the running coupling. 

The figure \ref{Vrun} shows an interesting result. The solid curve represents
the values of the potential evaluated by using the running coupling and the
dashed one the potential values calculated by employing constant coupling
values, chosen at the scale $\mu=11.63$ GeV. As it can be noticed the
substitution of the constant coupling by the running one, had not drastically
modified the minimum position at the mean field determining a Top mass value
near $175$ GeV. Therefore, a suspected beforehand possibility of a distortion
of the spontaneous symmetry braking pattern obtained at constant coupling, was
not realized: the pattern is mainly unaffected by the consideration that the
couplings varies with the momentum scale.

After evaluating the mass of the scalar field the result was close to the
corresponding outcome associated to a constant coupling
\begin{align}
\ m_{\phi}^{run}  &  =\sqrt{V_{run}^{total\text{ }\prime\prime}[0]}\nonumber\\
&  =46.5475\text{ \ GeV.}%
\end{align}

It should be remarked that the scalar field mass values obtained here are
smaller than the observed Higgs particle mass of $126$ GeV. After thinking
about this outcome, we consider that it does not represents a direct negative
result in connection of the studied possibility of basing the SM on a symmetry
breaking associated to the Yukawa Top-Higgs interaction. This conclusion is
determined by the following reasoning. If we consider a similar calculation of
the Higgs mass in the framework of the more complex SM, then, there will be
few more fields having similar values of mass as the Top quark (as the $W$ and
$Z$ bosons). Then, the contributions to the Higgs potential at two loops of
those modes can be expected to appreciably change the scalar field mass value
which could be determined only by considering the Top and the Higgs fields
(which are the only two fields included in the model). Therefore, the most
important outcome of the present work, should be considered as this one: to
determine that the spontaneous symmetry breaking generated by a single quark
and a scalar (upon fixing the observed quark top mass)  produces a mass for
the scalar field being smaller and close to the Higgs one. We estimate that
the results of the present work directly suggest the interest of attempting to
construct the SM upon the here investigated spontaneous symmetry effect.

\section*{Summary}

We have explored the possibility for that the spontaneous symmetry breaking
effect in the SM could be implemented thanks to the Yukawa interaction of the
Top quark with the Higgs field.  For this purpose a formerly proposed simple
model was reconsidered. The previous  work although indicating in some sense
the possibility investigated, was inconclusive due to the arising in it of a
small value of the renormalization scale (smaller than $1$ GeV ) in order to
allow the Top quark to get the observed value of its mass.

In the present work, we reevaluated the effective potential created by the
system for the scalar field, and found calculational errors. Their correction
then, led to a picture in which it is also possible to fix the Top quark mass
value, but at an intermediate value of the scale $\mu=11.63$ GeV. The value of
the scalar particle mass now emerging was of nearly $45$ GV which is close but
smaller than   the Higgs mass. However, being an amount of the same order
still allows for  the possibility that the $W$ and $Z$ contributions, to be
added in a calcualtion done  in  the framework of the SM, could rise the
result up to the observed Higgs mass of $126$ GeV.

The work also investigate the stability of the result for the spontaneously
symmetry pattern by considering the effect of employing  the running with
momentum coupling in the calculation. For this purpose,  the finite expression
in dimensional regularization of the quark-gluon contribution two loop
effective potential for the Higgs fields was  expressed as a momentum integral
through a specially designed substraction procedure. The difficult in
attaining this formula, was produced by the use of dimensional regularization.
In this scheme, the divergences are normally substracted after integrating
over the momenta. However, we required to conserve the momentum integral in
order to allow the substitution of the constant coupling by the running one.
Then a momentum integral was retained by firstly substracting to the momentum
integrand a relatively simple expression, which makes the momentum integral
finite. Afterwards, the integral that was substracted was exactly evaluated in
dimensional regularization. This allowed to determine the divergent pole part
and also the finite part which is also dependent of a parameter just
introduced for eliminating a remaining zero momentum divergence. The divergent
part just reproduced the minimal substraction counterterm of the quark-gluon
contribution to the two loop effective action. Finally the mentioned
additional parameter was fixed by imposing that the finite part of the
substracted integral vanishes for all values of the scalar field and scale parameter.

Further, the obtained formula for the potential (influenced by substituting
the running coupling) was calculated for the same value of scale parameter for
which the potential was before evaluated at a constant coupling. The formula
for the running coupling employed was a fit to the available data for the
measured couplings. The values of the coupling at small momenta were assumed
to be constant when the momentum value is smaller than the one associated to
the maximal value of the measured coupling. Surprisingly, the results for the
potential became very close to the ones evaluated for constant couplings at
the same scale. This outcome allows to conclude that the reduction of the
coupling with momenta does not disturb the arising spontaneous symmetry pattern.

In a future extension of this work we plan to start from a Lagrangian being
practically equivalent to the SM's one, in which all the Higgs field terms
associated to the usual scalar doublet will be present, but in which only the
negative mass squared term creating the Mexican Hat potential will not be
considered. The idea will be to attempt use the many parameters in this
slightly modified SM model, for implementing a symmetry breaking patterns
being similar to the one discussed here.

\acknowledgements A.C. would like to acknowledge a helpful discussion with
Prof. Masud Chaichian in which he underlined the motivating possibility of
checking whether the use of the running coupling constant in place of a
constant coupling would support \ the spontaneous symmetry breaking pattern.


\begin{thebibliography}{99}                                                                                               %


\bibitem {second-1}M. Gonderinger, Y. Li, H. Patel and M. Ramsey-Musolf, J.
High Energy Phys. 1001, 053 (2010)

\bibitem {second-2}J.A. Casas, J.R. Espinosa and M. Quiros, Phys. Lett. B 382,
374 (1996)

\bibitem {second-3}C.D. Froggatt, H.B. Nielsen, Phys. Lett. B 368, 96 (1996)

\bibitem {cabo}A. Cabo, Eur. Phys. J. C 71, 1620 (2011)

\bibitem {muta}T. Muta, \textit{Foundations of Quantum Chromodynamics}, World
Scientific Lectures Notes in Physics, vol. 5 (1987)

\bibitem {castano}H. Arason, D.J. Casta\~no, B. Kesthelyi, S. Mikaelian, E.J
Piard, P. Ramond and B.D. Wright, Phys. Rev. 46, 3945 (1992)

\bibitem {fleischer}J. Fleischer and O.V. Tarasov, Z. Phys. C 64, 413 (1994)

\bibitem {bateman}H. Bateman and A. Erdelyi, \textit{Higher Trascendental
Functions}, vol. 1, McGraw-Hill, New York (1953)

\bibitem {schloser}M.J. Schlosser, \textit{Multiple Hypergeometric
Series-Appell Series and beyond}, arXiv:1305.1966v1 (2013)

\bibitem {ParticleData}Review of Particle Physics, Journal of Physics G:
Nuclear and Particle Physics, Vol. 33 Pp. 1-1232 (2006)




















































\end{thebibliography}
\end{document}